\documentclass[11pt]{article}
\usepackage{blois,epsfig}

\def\mco{\multicolumn}

\def\ra{\rightarrow}

\def\ko{K^0}

\def\be{\begin{equation}}
\def\ee{\end{equation}}
\def\bea{\begin{eqnarray}}
\def\eea{\end{eqnarray}}

\begin{document}
\vspace*{4cm}
\title{  NEUTRINO MASS AND NEW PHYSICS}

\author{ A. YU. SMIRNOV~\footnote{Talk given at the XXI th Rencontres de 
Blois ``Windows on the Universe'', Cha¢teau Royal de Blois, 21st - 26th 
June, 2009}}

\address{International Centre for  Theoretical Physics, Strada Costiera 11,\\
34014 Trieste, Italy}

\address{Institute for Nuclear Research, Russian Academy of Sciences,\\
Moscow, Russia}

\maketitle\abstracts{
Implications of the recent results on neutrino masses and mixing  
to underlying new physics are considered. 
Various approaches to  physics behind neutrino mass 
are described which include the tri-bimaximal mixing 
and flavor symmetries, the 
quark-lepton complementarity and weak complementarity, 
the quark-lepton universality and unification. 
Some recent results from string phenomenology are discussed and 
the issue of model building versus ``string engineering'' 
is outlined.}

\section{Introduction}
%%%%%%%%%%%%%%%%%%%%%%%%%%%%%%%%%%%%%%%%%%%%%%%%%%%%%%%%

It is about  11 years since the  discovery of neutrino mass.   
Still, in spite of enormous efforts of many theoreticians 
and experimentalists the ``Physics behind neutrino mass''  
has not been identified. 
It should be some ``New physics''  beyond the Standard Model. 
It can be the old ``new physics'' invented many years ago 
and studied in details theoretically. 
It can be new ``New physics'' proposed recently,  
or something we have not thought about. 

The not-yet excluded new physics 
covers enormous range of possibilities: 
 
- from the eV  to Planck mass of the underlying scale 
(27 orders of magnitude); 

- from exact flavor symmetry to anarchy and randomness;  

- from quark-lepton unification to fundamental 
difference of quarks and leptons; 

- from attempts to explain the observed features 
in a single QFT context to idea that nothing can be explained 
in details and the observed features are result of 
complicated ``evolution'' from Planck or string scales to low energies.  
%%(like creation and evolution of the planetary system). 
The only what we really know is that all existing proposals 
can not be correct simultaneously.

The paper is organized as follows:
In Sec.~2 the existing information about masses and mixing
is presented and its immediate implications are discussed.
In Sec.~3 possible physics behind smallness of neutrino mass is
considered. Sec.~3 is devoted to different approaches to
explain lepton mixing.
Some recent results obtained in the string phenomenology
are described and the issue of whether the ends
(the bottom-up and top-down approaches) meet is addressed in Sec. 5.

\section{Data and Implications}
%%%%%%%%%%%%%%%%%%%%%%%%%%%%%%%%%%%%%%%%%%%%%%%%%

Analysis of results from neutrino  propagation experiments 
with neutrinos from the sun \cite{sun}, atmosphere~\cite{atm}, 
reactors~\cite{kamland} and accelerators~\cite{k2k,minos}  
in terms of oscillations~\cite{osc}
and adiabatic conversion (the MSW effect)~\cite{msw} gives now  
rather precise determinations of mass squared differences $\Delta 
m_{ij}^2$ and mixing angles $\theta_{ij}$. (For recent analysis see 
\cite{fogli,malt,valle,schwetz}.) Few conclusions follow immediately. 

1). The absolute mass scale: From MINOS \cite{minos} and atmospheric 
neutrino \cite{atm} data we have 
$ m_i > \sqrt{\Delta m_{31}^2} > 0.045~~{\rm  eV}$ ($i = 1,3$).
Cosmology~\cite{cosm} gives an  upper bound on the sum of masses which 
implies  
$m \approx  \frac{1}{3}\sum_i m_i < (0.2 - 0.3)~{\rm eV}$. 
Consequently, the heaviest neutrino should have 
the mass in the range 
\be
m =  (0.04 - 0.3)~ {\rm eV}, 
\ee
and the upper part of this region will be 
explored by KATRIN \cite{katrin}. 
%%which will start to operate in 2010. 

2). Mass hierarchy:  $\Delta m_{31}^2$ and 
small mass split, $\Delta m_{21}^2$, that follows from the solar neutrino  
experiments~\cite{sun} and KAMLAND~\cite{kamland}   
give
\be
\frac{m_2}{m_3} \geq \sqrt{\frac{\Delta m_{21}^2}{\Delta m_{31}^2}}
\approx 0.18, 
\ee
which means that neutrinos have the weakest mass hierarchy 
(if any) among the known fermions (quarks and leptons). The latter 
may be related to the large lepton mixing.

3). Nature of neutrino mass: The smallness  may indicate that 
nature  of neutrino mass (or at least what 
we extract from oscillation experiments) differs from the one of other 
fermions. Is $m_\nu$ of the same nature as the mass of electron or top 
quark?
Is
\be
m_\nu ({\rm oscillations}) = m_\nu ({\rm kinematics})?  
\ee
In general,  
\be
m_\nu = m_{standard} + m_{soft}(E,n), 
\ee
where $m_{soft}(E,n)$ is the 
medium (environment)  dependent (``soft'') component. 
Can $m_{soft}$ dominate? E.g. in \cite{pedro} it was shown that 
the density-dependent soft masses of the form   
$m_i = m_0  \tanh [\lambda_i \rho(g/cm^3)]$
with $m_0 = 5 \cdot 10^{-2}$ eV and $\lambda_i = (0,~ 0.06,~ 3)$ 
allow one to explain most of the oscillation data. 

Of course,  the outstanding dilemma is  Dirac versus  Majorana. 
%%is establish whether neutrino is Majorana Dirac vs Majorana
 
\begin{figure}
%%\rule{5cm}{0.2mm}\hfill\rule{5cm}{0.2mm}
%%\vskip 2.5cm
%%\rule{5cm}{0.2mm}\hfill\rule{5cm}{0.2mm}
\begin{center}
\psfig{figure=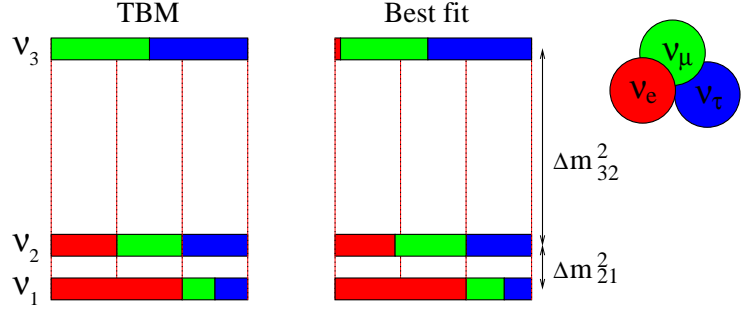,height=1.6in}
\end{center}
\caption{Neutrino mass and mixing spectrum which corresponds to 
the tri-bimaximal mixing (left) and the best fit values of the mixing 
angles (right).} 
\label{spectrum}
\end{figure}

4). Mixing: Information about mixing is encoded in the 
Fig.~\ref{spectrum}, where one can see few salient features: 
(i) admixture of the $\nu_e$ flavor in the third state 
is small or zero; (ii) the muon and tau flavors are mixed in this third 
state  almost equally; (iii)  all three flavors are mixed in the second 
flavor nearly equally: 
\be
|U_{e3}|^2 \equiv \sin^2 \theta_{13}  < 0.05; ~~~~
|U_{\mu 3}|^2 \approx |U_{\tau 3}|^2 \approx \frac{1}{2}; ~~~~ 
|U_{e2}|^2 \approx |U_{\mu 2}|^2 \approx |U_{\tau 2}|^2 \approx \frac{1}{3}. 
\label{trimax}
\ee
As a consequence,  
$\tan ^2 \theta_{12} = {|U_{e2}|^2 }/{|U_{e1}|^2} 
\approx {1/2}$, $\tan^2 \theta_{23} = 
|U_{\mu 3}|^2/|U_{\tau 3}|^2 \approx 1$. 
In the case  $|U_{e3}|^2  = 0$ and exact other equalities in 
(\ref{trimax})  we deal with 
the tri-bimaximal mixing (TBM)~\cite{tbm}. 
The latest experimental results, however, testify for certain deviation 
from the TBM-scheme.
In particular, complete three neutrino analysis of the 
atmospheric neutrino data, which includes the 1-2 mass 
split, leads to 
\be
\sin^2 \theta_{23} \sim 0.43 - 0.47 ~~~(0.5). 
\ee
(We show in brackets the TBM predictions.) 
Recent results from solar neutrinos, in particular,  
lower than before ratio of the charged to neutral current rate,  
give $\sin^2 \theta_{12} \approx 0.31 ~~(0.33)$.
There are certain indications that the 1-3 mixing is nonzero 
and actually not very small. 
According to the global analysis of the oscillation data 
in~\cite{fogli}
\be
\sin^2 \theta_{13} \approx 0.016 \pm 0.010. 
\label{13nonz}
\ee
MINOS ~\cite{minos-e}
has found some excess of the $e-$like events which 
can be interpreted as due to non-zero 1-3 mixing. 
Adding this to (\ref{13nonz}) leads to  
\be
\sin^2 \theta_{13} \approx 0.02 \pm 0.01 ~~(1 \sigma). 
\label{13nonzt}
\ee
An independent analysis~\cite{malt} shows essentially no hint from 
atmospheric neutrino data.  
Exact TBM  agrees with data at about $(2 - 3) \sigma$ level. 
%%%%%%%%%%%%%%%%%%%%ffff1%%%%%%%%%%%%%%%%%%%%%%%%%%%%%%%%%%%%%%%%%%%%
%%%%%%%%%%%%%%%%%%%%%%%%%%%%%%%%%%%%%%%%%%%%%%%%%%%%%%%%%%%%%%%%%%
Are the deviations from TBM small? 
In terms of $\sin^2 \theta_{13}$  they indeed look small.  
However, from theoretical point 
of view $\sin \theta_{13}$ is more appropriate  
for which we obtain from Eq. (\ref{13nonzt}) 
0.15 (to be compared  
with  $\sin \theta_{12} \sim 0.55$).  
%%Still, in contrast to quark sector 
%%$\theta_{12} \cdot \theta_{23} > 2 \theta_{13}$. 

5). One clear conclusion is that the 
patterns of lepton and quark mixings are strongly different. 

The same data considered in terms of mixing angles 
favor another interpretation: 
comparing quark and lepton mixings one finds that 
the 1-2 and 2-3 mixing angles  in quark 
and lepton sector sum up to maximal mixing angle:  
\be
\theta^l_{12}  +  \kappa \theta^q_{12} \approx \pi/4, ~~~
\theta^l_{23}  +  \kappa \theta^q_{23} \approx \pi/4
\label{angleqlc}
\ee
with  $\kappa = 1/\sqrt{2}$ or 1 depending on the way of summation  
\cite{qlc}.\\  
%%1-3 mixing is small in both sectors. \\

What can be inferred from the data? 
Few observations are in order. 
(i) The data show both order, regularities 
and some degree of randomness. 
(ii) No simple relations between all masses and mixings 
have been found which could testify for simple 
underlying scenario. 
(iii) Different pieces of data indicate different 
underlying physics. Indeed, the scale of neutrino masses may testify for 

- Grand Unification (quark-lepton correspondence) 
via seesaw;  

-  extra dimensions,  especially 
if  masses are of the Dirac type;  

-  certain symmetry which suppresses  
the electroweak scale for neutrino mass; 

-  absence of the RH neutrino components.\\  
\noindent
The pattern of lepton mixing  
indicates two completely different possibilities:  

- existence of flavor symmetries 
(in addition, this is supported by fermion mass hierarchy, 
and Koide relations~\cite{koide} for the masses of charged leptons); 

- ``anarchy'' \cite{anarchy}.

\section{Mass scales and mechanisms}
%%%%%%%%%%%%%%%%%%%%%%%%%%%%%%%%%%%%%%%%%%%%%%%%%%%%%%%%%%%

There are two issues related to the scale of neutrino mass: 
(i) suppression of ``natural'' electroweak scale; 
(ii) generation of small mass measured in oscillations. 
Seesaw mechanism~\cite{type1} does these two things simultaneously.  
No additional symmetry is needed. 
On the other hand, the seesaw can only work as a mechanism of suppression
if, e.g., the RH neutrino masses are at the Planck scale. 
Then dominant contribution to neutrino mass  
comes from some other mechanism. Apart from the seesaw,  suppression 
of the EW scale mass can be due to 
certain symmetry (which looks  unnatural if it is introduced for the 
RH neutrinos only), 
or multi-singlet mechanisms which rely on existence of new neutral 
leptons, singlets - of the SM symmetry group.  

There are  several different versions of the seesaw  mechanism.  
If at the effective level neutrino masses 
are generated by the D=5 operator with only SM fields,  
$LLHH/M$~\cite{weinberg} , there are three tree level realizations  
related to the type of exchanged heavy particles:  
type-I (with exchange of singlet RH neutrinos) ~\cite{type1}; 
type-II (with $SU(2)$-triplet scalar boson)~\cite{type2},  
type-III \cite{type3} (with neutral fermion 
from the $SU(2)$ triplet).  

One can invent other mechanisms which lead  
at the effective level to the D=5 operators 
with  higher representations for the 
scalar fields $\Phi_k$ ($k$ being the dimension of representation):
$M^{-1} L L \Phi_k \Phi_n$. 
Also high dimension operators can be used with more than two 
scalar fields which would require lower scale of the underlying physics
(see, e.g. \cite{Babu:2009}). 

The extra (spatial) dimension mechanisms are based 
on the overlap suppression and natural for the Dirac neutrinos.
Indeed, different localizations of the 
LH the RH neutrino components can be related to 
their different gauge properties ($N_R$ 
have no SM interactions).  
The Yukawa coupling is proportional to 
degree of overlap of the LH and RH wave functions $\alpha$,  
so that  in 4D we have
\be
m_{EW} \alpha \bar{f}_L f_R + h.c..  
\ee 
Different versions of this suppression 
depend on an  extra-dimensions setup.  
In the ADD case (flat extra dimensions)~\cite{add}, 
$N^c$ being a singlet of SM symmetry group can reside 
in the bulk of extra dimensions, whereas 
$\nu_L$ is localized on the brane, 
so the overlap equals $\alpha = M_*/M_{Pl}$ where $M_*$ 
is the fundamental scale of theory. 
In the RS scenario, 
$\nu_L$ and $N^c$ are localized on  different branes 
(the EW and Planck ones)~\cite{rs} with exponentially 
decreasing wave functions  with distance 
from the corresponding branes. Here 
$\alpha \sim  M_{Pl} (v_{EW}/M_{PL})^{\nu + 0.5}$, 
where $\nu = 1.1 - 1.6$. 

Small effective Yukawa couplings can be a consequence 
of certain symmetry which suppresses the couplings at the renormalizable 
level. Then non-renormalizable operators
\be 
a \bar{L} \nu_R H \frac{S}{M}
\ee 
produce effective  Yukawa coupling 
$ h = a \langle S \rangle / M $. 
For $a = O(1)$,  we need  $\langle S \rangle /M \sim 10^{-13}$. 
This smallness can appear e.g. as the ratio 
of  SUSY and GUT (or Planck) scales: 
$h \sim m_{3/2}/ M_{Pl}$,  etc.. 
One can consider higher dimensional operators
\be 
a \bar{L} \nu_R H \frac{S_1 ...S_n}{M^n}. 
\ee

Supersymmetry  opens various new possibilities: 
there are genuine SUSY mechanisms 
of neutrino mass generation 
with rather particular features. 
SUSY provides new mass scales: 
the SUSY breaking scale, $m_{3/2}$, 
as well as the SUSY conserving mass term 
for two Higgses $\mu$ \cite{kitano}. The neutrino  mass can appear as 
\be
m_\nu \sim \frac{1}{M_{GUT}} \mu v_{EW}~~ {\rm or }~~ 
m_\nu \sim \frac{1}{M_{GUT}} m_{3/2} v_{EW} . 
\ee 
In SUSY neutrinos are not unique:  
the neutralinos, have similar properties. 
Neutrinos can mix with neutralinos
(if R-parity is broken) which leads to yet another mechanism of neutrino 
mass generation \cite{rpar}. 

\section{Mixing and new physics} 
%%%%%%%%%%%%%%%%%%%%%%%%%%%%%%%%%%%%%%%%%%%%%%%%%%%%%%%%%%

There are several approaches  
to understand neutrino mixing which have  different 
implications for the fundamental theory.  
They differ by possible relations between masses and mixings 
and also by degree of connection of quarks and leptons.

\subsection{Tri-bimaximal mixing}
%%%%%%%%%%%%%%%%%%%%%%%%%%%%%%%%%%%%%%%%%%

\begin{itemize}

\item
It is assumed that the  (approximate) 
TBM is not accidental but  
rather straightforward consequence of some flavor 
symmetry $G_f$. 

\item
This implies the form-invariant mass matrices and 
absence of relation between masses and mixing. 
Extensions to quarks are usually problematic.

\end{itemize}

The symmetry $G_f$ should be broken spontaneously or 
be valid  in some  part of the Lagrangian only. 
The most popular and minimal 
(?) 
%%(see Lam vs. Grimus)
is $G_f = A_4$ \cite{ma} (see general statements in \cite{lam,grimus}). 
The group has triplet (real) representation 
and three inequivalent singlet representations. This is somehow minimal 
set which gives enough freedom to construct viable models.  
Other  possibilities include $G_f = $ $S_4$,  
$T_7$,  $\Delta(3 n^2)$, etc..

General idea to explain mixing is the following. 
Mixing appears as a result 
of different ways of the original 
flavor symmetry  breaking in the neutrino and charged lepton 
sectors. Symmetry is not broken completely: 
there are certain residual symmetries in both  
(Yukawa) sectors: 
\be
G_f \rightarrow 
\left\{
\begin{array}{ll}
G_l & ~~ {\rm for ~charged ~leptons}\\ 
G_\nu & ~~{\rm for~ neutrinos} . 
\end{array}
\right. 
\ee
Furthermore,   the  residual symmetries are different:  
$G_l \neq G_\nu$.  These symmetries determine 
certain structures of the mass matrices
which then lead to the required mixing: 
E.g. $G_l$ ensures that the charged lepton mass matrix 
$M_l$ is diagonal, whereas the neutrino mass matrix 
$M_\nu$ is of the TBM-type. 
Usually  
(at least in the case of $A_4$) $G_\nu$
is not enough to fix the TBM form 
and some additional symmetry, like 
$\nu_{\mu} \leftrightarrow \nu_{\tau}$,  
$A_{\mu\tau}$, is needed. It can appear as an  
``accidental'' symmetry which is a consequence   
of specific choice of the flavon representations and particular 
configuration of VEV's. 

In turn, different ways of the symmetry breaking 
in the neutrino and charged lepton sectors originate  
from different flavor (symmetry) assignments  for  
the right handed  components of leptons: $N^c$ and $l^c$, 
and correspondingly,  different Higgs multiplets which 
give masses to charged leptons and neutrinos.  
Thus, origins of mixing are in different 
symmetry assignments for  the RH components. 
In many cases this is inconsistent with 
the  L-R symmetry and grand unification.  
At the same time, the RH components have 
different EW hypercharges. So,  one can somehow 
correlate different Yukawa sectors with hypercharges. 
All this looks rather complicated but, in fact,  
something similar may follow from string theory (see below).  
The difference of mixings can also be related to 
existence of the Majorana mass terms of RH 
neutrinos (whereas the Dirac 
sectors could be similar in quark and lepton sectors).  

\begin{figure}
%%\rule{5cm}{0.2mm}\hfill\rule{7cm}{0.2mm}
%%\vskip 2.5cm
%%\rule{5cm}{0.2mm}\hfill\rule{5cm}{0.2mm}
\begin{center}
\psfig{figure=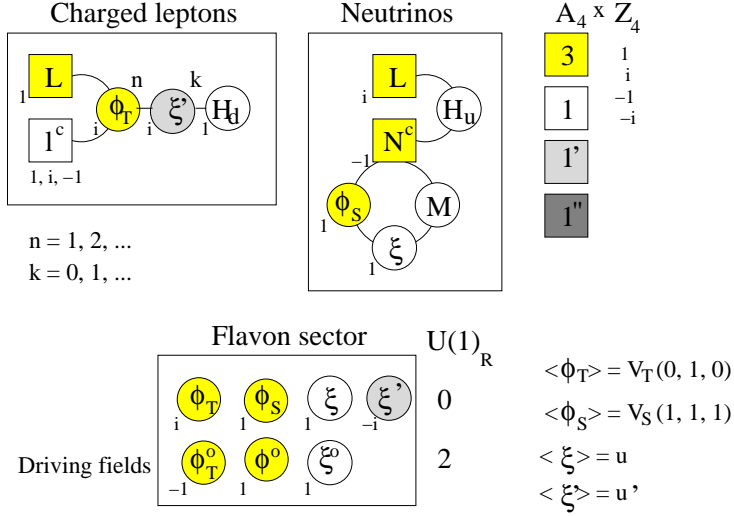,height=2.7in}
\end{center}
\caption{The Yukawa sector and flavon content of the model$^{~32}$.
%%~\cite{altarelli}. 
Colors indicate transformation properties of the corresponding 
multiplets with respect to $A_4$; complex numbers at the notations 
of multiplets give the transformation properties with respect to $Z_4$.}
\label{a4model}
\end{figure}

Let us illustrate realization of this idea 
using model \cite{altarelli}, which is probably the simplest 
and the most advanced, in a sense that most of the features 
are explained using symmetry (choice of field multiplets and 
symmetry assignments).  
The model is based on the $A_4 \times Z_4$ flavor group. 
Mixing originates both from different symmetry properties of  
the RH components of charged leptons and neutrinos and 
from the Majorana mass terms of the RH neutrinos.  
A general structure of the Yukawa sector of the 
model is shown in Fig.~\ref{a4model}.

%%%%%%%%%%%%%%%%%%%%ffff2%%%%%%%%%%%%%%%%%%%%%%%%%%%%%%%%%%%%%%%%%%%%
%%%%%%%%%%%%%%%%%%%%%%%%%%%%%%%%%%%%%%%%%%%%%%%%%%%%%%%%%%%%%%%%%%

Indeed,  $l^c$  are singlets $1_1$ of $A_4$ and transform as 
with $1, i, -1$ under $Z_4$ transformations. 
In contrast, $N^c$ form triplet of $A_4$ and all three components 
transform with $-1$ under $Z_4$.  
Higgs sectors for neutrinos and charged leptons are not symmetric. 
Flavor symmetry is broken by flavons. 
The charged lepton masses are generated 
by non-renormalizable terms
\be
\frac{1}{\Lambda_f^{n + k}} L l^c (\phi_T)^n (\xi^\prime)^k H_d.
\ee
In contrast, the neutrino Dirac mass    
is generated without violation of flavor at three level:  
$
L N^c H_u. 
$
The  Majorana masses have both 
flavor conserving and flavor violating contributions: 
\be
N^c N^c (M + g \xi) +   N^c N^c \phi_S
\ee
($g$ is constant). Notice that  flavons involved here are  different 
from those for charge leptons. 

The symmetry $Z_4$  (in some other versions   
U(1)) is required to produce hierarchy of 
charged lepton masses as well as to forbid ``unwanted'' interactions. 

Some comments which have generic character for this model building 
follow.

1). {\it Symmetry properties assignment} is done essentially 
{\it ad hoc} 
which should be considered as  another ``discrete''  degree of freedom. 
According to Fig.~\ref{a4model} the assignment looks rather accidental  
(there is no guide or system or rule),  some 
simple representations are missed.  
It is rather doubtful that this can be embedded
in an extended group structure.
The assignment prevents from immediate  
grand unification like SO(10) 
or even L-R symmetry, and further substantial 
complications are required. 

2). {\it Vacuum alignment of flavons.} 
Flavor symmetry should be broken (see however, \cite{grimus2} 
for scenario with unbroken or softly broken symmetry). 
Therefore the observed  flavor structure depends both on 
symmetry of couplings and  also on vacuum configuration. 
%%Vacuum alignment is actually a half of the program...
To achieve the required alignment one needs 
to further extend the model: 
introduce ``driving'' scalar fields (flavons), 
and to use supersymmetry.  
(Flavons and driving flavons differ by $R-$charges  
with respect to $U(1)_R$ symmetry). 
Still, masses of neutrinos are not defined
and only some bounds exist. 

An alternative would be to employ physics of extra dimensions. 
In fact, extra D offer a different origin of 3 generations 
\cite{frere}. That changes whole approach to fermionic masses. 

%%The TBM requires certain symmetry breaking. 

Complexity of models and deviations of angles from TBM values add 
some more doubts in correctness of interpretation of 
the observed tri-bimaximal mixing. Still several possibilities exist: 
(i) TBM is exact  and it implies 
existence of underlying flavor symmetry;  
(ii) TBM  is only the first approximation 
and deviations exist which are consequences of the 
flavor symmetry violation and RGE effects;  
(iii) the approximate TBM is just   
numerical coincidence without any fundamental 
implications.  Immediate flavor symmetry is  
misinterpretation. TBM may be a result of 
interplay of different factors not related to symmetry;  
(iv) large deviations from TBM are possible and 
the observed mixing has origins unrelated to the flavor symmetries. 

\subsection{Quark-lepton complementarity}
%%%%%%%%%%%%%%%%%%%%%%%%%%%%%%%%%%%%%%%%%%%%%%%%%

This concept is based on observations in Eq. (\ref{angleqlc}). 
Qualitatively one finds certain correlation of angles: 

the 2-3 leptonic mixing is close to maximal
because the 2-3 quark mixing is small; 

the 1-2 leptonic mixing deviates from 
maximal substantially because 
the 1-2 quark mixing is relatively large.

In other words
${\rm ``lepton~ mixing = bi-maximal~ mixing - quark~ mixing''}$. 
Possible implications of QLC~\cite{qlc}: 

\begin{itemize}

\item
Quark - lepton symmetry or unification, 
or alternatively, common flavor symmetry.  

\item
Existence of structure in theory  which generates 
the bi-maximal mixing. 
%%Seesaw itself can be  responsible for this. 
%%In this way smallness of neutrino mass is related to 
%%the bi-maximal mixing. 
%%[[Zee mechanism]] 

\end{itemize}

One can develop some perturbative realization of QLC.  
In the lowest order   
the quark mixing is absent and the lepton mixing is the bi-maximal 
$V_{BM} = V_{23} V_{12}$ (product of the two rotations on the angles 
$\pi/4$): 
%%(also masses of fermions from two generations are zero):  
\be
U_{CKM}^0 = I, ~~  U_{PMNS}^0 = V_{bm}. 
%%m_1 = m_2 = 0. 
\label{ckm0}
\ee
The CKM mixing, the  deviations of lepton mixing from 
bi-maximal and possibly generation of light fermion  masses have the same 
origin. 
%%[[model? see Altarelli]]
In the context of seesaw certain structure of 
the RH neutrino mass matrix can produce the bi-maximal mixing. 
The Dirac mass matrices  (related by GUT) are origins of CKM 
and the deviations.

The deviations 
%%from the leptonic bi-maximal mixing 
%%and unit quark mixing, as well as masses of light generations 
may not be (at least directly) related  to GUT or quark-lepton symmetry, 
but be a generic features of flavor physics. Indeed, 
(i) QLC relations are not exact (though the deviations can be well 
due to RGE effect \cite{schmi} or some other corrections). 
(ii) There are relations of the type  
\be 
\sqrt{\frac{m_d}{m_s}} \sim ~
\sqrt{\frac{m_\mu}{m_\tau}} \sim \theta_c.  
\ee
That is, the same parameter of the  Cabibbo angle size  appears  
in various places,  
and it can be considered as a kind of ``quantum'' of flavor 
physics \cite{qlc,Rodejohann}.  This feature was called ``Cabibbo haze'' 
in Ref.~\cite{haze} and recently, the ``weak complementarity'' 
in Ref.~\cite{weakc}.

Again it  may have some perturbative realization:  
the deviations are due to 
the high order corrections. In models with flavor symmetry and 
flavons the lowest order generates mixing given in Eq. (\ref{ckm0}). 
The first order corrections   $\sim  \langle \phi \rangle /\Lambda_f$ 
generate simultaneously the Cabibbo mixing and 
deviation from bi-maximal mixing: GUT is not necessary. 
Due to residual or accidental symmetry some corrections 
may appear in higher orders, thus producing hierarchies 
of three generations. Model of this type has been 
proposed in \cite{alt2}.

\subsection{Quark-lepton universality and unification}
%%%%%%%%%%%%%%%%%%%%%%%%%%%%%%%%%%%%%%%%%%%%%%%%%%%%%%%%%

In this approach there is no fundamental difference 
between quarks and leptons. 
No special symmetries in the quark and/or lepton sectors exists.  

\begin{itemize}
\item
There is  the quark-lepton  symmetry, or correspondence, 
or unification  with SO(10) being the most appealing.  

\item
The differences between quark and lepton mixings and mass spectra 
originate from  differences of known (gauge) properties 
of quarks and leptons, in particular, neutrality of neutrinos. 

\end{itemize}

The arguments in favor of this point are   
(i) apparent quark-lepton correspondence
which can be described in terms of 
the Pati-Salam symmetry - consideration of leptons as 4th color;  
(ii) embedding of known fermions plus 
well motivated RH neutrino into 16-plet of SO(10). 
(iii) unification of three different interaction. 
It is difficult to believe that these facts are  accidental.  

Smallness of neutrino mass itself can testify for GUT. 
In the context of seesaw, the required values of masses of the RH neutrinos 
indicate existence of mass scale which is close or coincides with 
the GUT scale.  There are two possibilities here.  
In the presence of mixing the largest  RH neutrino mass 
can simply coincide with the GUT scale: 
$M_R \approx M_{GUT}$. In this case masses should have strong 
(quadratic) hierarchy $M_2 \sim (10^{12} - 10^{13})$ GeV 
and $M_1 \sim 10^{8}$ GeV, which in turn, can lead to the large lepton mixing  
via the seesaw enhancement \cite{seesenh}.  
The CP-violating decays of $N_1$ can realize 
the leptogenesis in the Universe~\cite{leptogen}.   
Another possibility: 
$M_R \approx M_{GUT}^2/ M_{Pl}$. It can be realized 
if  singlet fermions with masses at the Planck scale exist and 
mix with RH neutrinos at $M_{GUT}$ (double seesaw)~\cite{doubless}.  
 
The difference of mass and mixing spectra of 
quarks and leptons can be related to the neutrality of neutrinos.   
There are various realizations of this idea. 

1. Seesaw itself: due to neutrality the RH neutrinos have large Majorana 
masses, which leads to smallness of neutrino mass,  and 
simultaneously, particular structure of the RH 
neutrino mass matrix enhances  mixing.  

2. Singlet fermions  (from hidden sector) can exist.   
Due to neutrality only neutrinos can mix with these singlets. 
Possible scenario is, e.g. the  SO(10) model with 16 plets 
of fermions plus some number (which, in fact, can be as big  
as several hundreds) of singlet fermion fields~\cite{hagedorn}. 
These singlets can mix both with usual LH  
and RH neutrino components. Such a mixing 
can (i) decrease the effective scale of seesaw; 
(ii) enhance mixing; (iii) produce zero order mixing; 
(iv) ``screen'' the  Dirac mass hierarchies in the 
context of seesaw; (v) produce randomness (anarchy) in  
the light neutrino mass matrix, and
consequently,  lepton mixing;  
(vi) explain smallness of neutrino mass; 
(vii) generate accidental seesaw symmetries.

\section{Do the ends meet?}
%%%%%%%%%%%%%%%%%%%%%%%%%%%%%%%%%%%%%%%%%%%%%%%%%%%5

In the previous sections we have described attempts to 
identify new  underlying physics starting from the data.   
Does it matches with what string theory can offer? 
String theory is expected to provide a 
guidline and context 
which should help to select among different bottom-up scenarios.  
It can  indicate that something is missed in our QFT considerations.

In general, string theory offers the following ``menu'' for 
the effective field theory: 
(i) GUT, (ii) existence of a number O(100) of singlets of the 
Standard Model as 
well GUT symmetry group,  (iii) several U(1) gauge factor;  
(iv) existence of discrete symmetries;
(v) heavy vector-like families;
(vi) various non-renormalizable interactions  
 (with Planck/string cutoff scale); 
(vii) explicit violation of symmetries;  
and extradimensional mechanisms of symmetry breaking;  
(viii) incomplete GUT multiplets,  etc.. 

Do results of bottom-up approach  match with what string theory can offer? 
Do the ends meet?
Many of these elements have been already employed 
in the model building. String engineering versus model building 
is the key issue.

The engineering means essentially a play with geometry of 
the internal space and here we outline one interesting possibility. 
The F-theory whose phenomenology was elaborated recently \cite{vafa}  
can give important insight into underlying physics.    
The F-theory is certain version of the strongly coupled type IIB string 
theory. Compactification is ``engineered'' in such a way that 
F-theory leads to N = 1 supersymmetry. The corresponding 
internal space (bulk) where gravity  
propagates is a complex threefold $B_3$ (6 real dimensions). 
So, the staring point is 10D field theory. 

The gauge field degrees of freedom 
reside on 8 dimensional surfaces, $S$, 
$S^\prime$, etc., embedded in 10D bulk and wrapped 
by seven-branes (the internal dimension of $S$ is 4).  
Different gauge groups are associated with different surfaces. 
The GUT group is engineered to be $SU(5)$ or of higher rank.  

In F-theory an additional condition is imposed that gravity 
decouples from the GUT. 
Decoupling implies that theory does not contain 
adjoint chiral superfields to break SU(5). 
Therefore breaking of F-theory SU(5) (with the decoupling) 
requires introduction of a flux in the 
hypercharge direction $U_Y(1)$. This leads to 
GUT symmetry breaking down to the SM symmetry group and 
also to important consequences for the Yukawa couplings.  

\begin{figure}
%%\rule{5cm}{0.2mm}\hfill\rule{7cm}{0.2mm}
%%\vskip 2.5cm
%%\rule{5cm}{0.2mm}\hfill\rule{5cm}{0.2mm}
\begin{center}
\psfig{figure=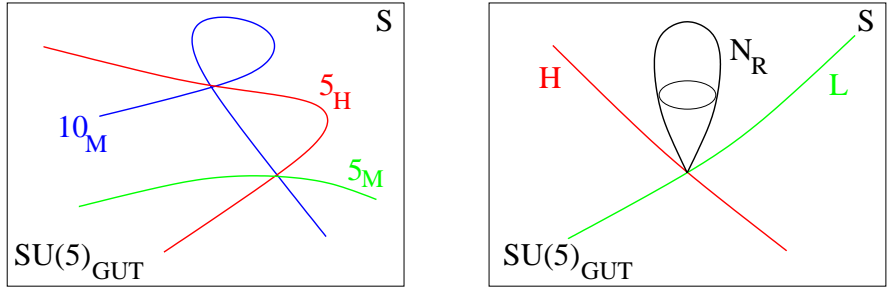,height=1.5in}
\end{center}
\caption{Generation of the Yukawa couplings in the F-theory 
for quarks and charged leptons (left) and neutrinos via 
the KK-seesaw (right). From Ref.$^{~45}$.
%%~\cite{vafa}.
}
\label{intspace}
\end{figure}

The matter and Higgs fields are localized on complex curves 
in $S$ (2 real internal dimensions). These curves are formed by 
intersections of the surfaces $S$, $S^\prime$,  etc., 
(see Fig.~\ref{intspace} from~\cite{vafa}). 
The Yukawa interactions appear at the intersections of these 
complex curves (real planes).  

%%%%%%%%%%%%%%%%%%%%ffff3%%%%%%%%%%%%%%%%%%%%%%%%%%%%%%%%%%%%%%%%%%%%
%%%%%%%%%%%%%%%%%%%%%%%%%%%%%%%%%%%%%%%%%%%%%%%%%%%%%%%%%%%%%%%%%%

The  main features of the fermion masses and mixing in  
this context can be formulated in the following way. 

1. In the lowest order the Yukawa couplings are given by  
the overlap of the 6D fields localized on 
the ``matter curves''. They appear  
at intersection of three matter curves which correspond to 
matter and Higgs fields. 
In the case of single intersection of lines of a given type, 
singular Yukawa matrices are generated 
\be
Y_{ij} \sim z_i z_j. 
\ee
Consequently, only one eigenvalue (mass) is non-zero for each type 
of fermions.
  
2. Masses of lighter quarks and leptons and mixings appear 
as a result of corrections 
due to interactions of matter fields  with the background gauge fields 
(hyperflux).
Consequently, the  corrections are determined by the gauge coupling.  
Indeed, the corrections are given by 
\be
\epsilon \sim \frac{1}{(M_* R_i)^2}, 
\ee
where $M_*$ is the compactification scale  of F-theory  and 
$R_i \sim 1/M_{GUT}$  are the lengths of the matter curves. 
So,  
$
\epsilon \sim ({M_{GUT}}/{M_*})^2. 
$
Using relation between 
the GUT and string (compactification) scale: 
\be
M_*^4 = \alpha_{GUT}^{-1} M_{GUT}^4
\ee
we obtain 
\be
\epsilon \sim \sqrt{\alpha_{GUT}}. 
\label{eps-def}
\ee
Mass matrix elements  appear then as powers of this parameter. 
 
3. The fact that GUT symmetry is broken in the hypercharge direction 
implies that expansion parameters for  
fermions with different hypercharges are different. 
So,  the origin of Yukawa structures (hierarchies) 
is in the gauge sector.  

4. Large lepton mixing is related to weak mass hierarchy 
of neutrinos and originates from particular 
properties of RH neutrinos or 
objects which play the role of the RH neutrinos in the F-theory context. 
Notice that the lines of $N_R$ are not in the $SU(5)$  
surface but they can cross this surface in one point. 
This point can be rather far from the intersection of 
$L$ and $H$ curves which can be used to 
achieve smallness of the Yukawa coupling.  

One interesting possibility is the Kaluza-Klein seesaw. 
The KK-mass term has the Dirac form $N_R N_R^c$, where 
only $N_R$ couples with usual lepton doublet and Higgs.
Therefore the starting point is structure with covering symmetry and 
two different sectors with couplings $\bar{L} N_R H$ and 
$\bar{L}^\prime N_R^c H^\prime$.  
Then identification $N_R \leftrightarrow N_R^c$, 
$L \leftrightarrow L^\prime$, etc.. 
(see Fig.~\ref{intspace} right) leads after integration out  
of the RH neutrinos to the effective 5D operator 
\be
m = \frac{1}{\Lambda_{UV}} L H_u L H_u . 
\ee
The fact that infinite tower of states contributes to neutrino mass  
explains an enhanced mixing and softer mass hierarchy.  
Up to O(1)  coefficients the lepton mixing matrix 
has the form \cite{vafa}
\be
U_{PMNS} \sim  
\left(
\begin{array}{ccc}
1 & \epsilon^{1/2} & \epsilon  \\
\epsilon^{1/2}  & 1 & \epsilon^{1/2}  \\
\epsilon & \epsilon^{1/2}  & 1 
\end{array}\right). 
\ee
Therefore $\theta_{12} \sim \theta_{23} \sim \alpha_{GUT}^{1/4}$, 
$\theta_{13} \sim \alpha_{GUT}^{1/2}$. 
In quark sector $\sin \theta_C \sim \sqrt{\alpha_{GUT}}$. 

The proposed F-theory scenario may have a perturbative 
field theory interpretation. This consideration 
is closer to the  third (unification) 
bottom-up approach. The model obtained 
should include additional heavy fields and symmetries and 
looks rather complicated and artificial~\cite{randall}. 
Structure of the mass and mixing matrices can be 
described by the $U(1)$  Peccei-Quinn symmetry with  expansion parameter 
$\epsilon$. This to a large extend reproduces features 
of the Froggatt-Nielsen mechanism. 
Although here  
the expansion parameter, the  Cabibbo angle,  is determined 
by the gauge  coupling.
Some discrete symmetries, like $S_4$, can appear as a result of symmetry 
with respect to rotations in the high dimensions.

\section{Conclusions}
%%%%%%%%%%%%%%%%%%%%%%%%%%%%%%%%%%%%%%%%%%%%%%%%%%%%%%%%%

What is new physics behind neutrino mass? No quick and simple answer, 
no unique and convincing scenario.  
Neutrinos did not yet help us to resolve  flavor problem, 
but added new puzzle. 

Approximate tri-bimaximal mixing  or 
quark-lepton complementarity  -
appealing phenomenological schemes. 
Still they can be accidental and misleading, without any 
fundamental implications.   
Plausible scenario of the underlying physics could include  
(i) see-saw  with high mass scales, 
(ii) (broken) flavor symmetry,  
(iii) quark-lepton symmetry, unification.

Further real progress requires 
checks of the proposed  ideas. This will be done to some extend 
with  forthcoming results on lepton number violating  decays, 
(MEG started to release the first results \cite{meg}), 
measurements of leptonic EDM,  
results from B-physics, proton decay searches, 
LHC and other high energy  accelerator experiments,  
Cosmology and  leptogenesis. 
Precision studies of neutrinos may reveal some new features.

Identification of the correct scenario will require 
further phenomenological and experimental developments, and probably 
long way of exclusion of different possibilities. 
LHC and other high energy experiments may shed some more light 
on to the problem  and probably identify correct context. 

\section*{Acknowledgments}

I would like to than J. Tran Thanh Van for invitation 
to give this talk. 

%%\section*{Appendix}
%%%%%%%%%%%%%%%%%%%%%%%%%%%%%%%%%%%%%%%%%%%%%%%%%%%%%%%%%%%%%

%%We can insert an appendix here and place equations so that they are
%%given numbers such as Eq.~\ref{eq:app}.

\section*{References}
%%%%%%%%%%%%%%%%%%%%%%%%%%%%%%%%%%%%%%%%%%%%%%%%%%%%%%%%%%

\end{document}